\documentclass[twocolumn,showpacs,preprintnumbers,amsmath,amssymb]{revtex4}
\usepackage{graphicx}
\usepackage{dcolumn}
\usepackage{bm}
\usepackage[utf8]{inputenc}
\usepackage{bm}
\usepackage{textcomp}
\usepackage{amsfonts}%

\begin{document}


\title{Dynamics of driven granular suspensions}


\author{Andrea Fiege}
\affiliation{Georg-August-Universität Göttingen, Institut für
  Theoretische Physik, Friedrich-Hund-Platz 1, 37077 Göttingen }
\author{Annette Zippelius}
\affiliation{Georg-August-Universität Göttingen, Institut für
  Theoretische Physik, Friedrich-Hund-Platz 1, 37077 Göttingen}
\affiliation{Max-Planck-Institut für Dynamik und Selbstorganisation,
  Am Fa\ss berg 17, 37077 Göttingen}
\affiliation{}


\date{\today}

\begin{abstract}

  We suggest a simple model for the dynamics of granular particles in
  suspension which is suitable for an event driven algorithm, allowing
  to simulate $N=\mathcal{O}(10^6)$ particles or more. As a
  first application we consider a dense granular packing which is
  fluidized by an upward stream of liquid, i.e. a fluidized bed. In
  the stationary state, when all forces balance, we always observe a
  well defined interface whose width is approximately independent of
  packing fraction. We also study the dynamics of expansion and
  sedimentation after a sudden change in flow rate giving rise to a
  change in stationary packing fraction and determine the timescale to
  reach a stationary state.


\end{abstract}


\maketitle

\section{Introduction}

Fluidized beds are ubiquitous in chemical and pharmaceutical
processes, including tablet coating, catalytic cracking, coal
combustion and incineration \cite{kunii1991fluidization}. They consist
of a collection of solid particles immersed in a fluid which can be a
gas or a liquid. The fluid is flowing upwards through the particles
and fluidizes the packing. Despite the simple setup, fluidized beds
show a variety of complex flow regimes. Gas-fluidized beds usually are
unstable and exhibit bubbling instabilities
\cite{davidson1971fluidization}, however intervals of stable
fluidization have also been
observed~\cite{geldart1986gas,tsinontides1993mechanics}
Liquid-fluidized beds are more stable and a wider range of flowrates
gives rise to homogeneous fluidization~\cite{Guazzelli1}, in
particular for low Reynolds number flow. Instabilities have been
observed in the form of voidage waves. These waves can be
one-dimensional in narrow beds \cite{anderson1969fluid} or develop
transverse structures in wider beds \cite{el1976instability}, that
have been conjectured to cause bubbles \cite{batchelor1991formation}.


Despite their seemingly simple ingredients, fluidized beds are far
from being well understood. As fluidized beds consist of particles and
a surrounding fluid,  both, interparticle- and particle-fluid-interactions
must be accounted for in an appropriate model.  Modeling the
fluidized bed by multiphase continua has been proven successful to
describe the onset and propagation of instabilities
\cite{jackson1988,jackson2000dynamics}. Stronger simplifications
were adopted by Johri and Glasser
\cite{johri2002connections,johri2004bifurcation}, who showed that a
suspension with nonuniform concentration can behave like a continuum
with nonuniform density subject to a density-dependent force. Although
this is a major simplification, the model is able to predict wavelike
instabilities in narrow fluidized beds.

Homogeneous fluidized beds have proven useful to study granular matter
near the jamming transition~\cite{Schroeter4,Schroeter2}. Fluidization
allows to adjust the volume fraction, so that e.g. the onset of
mechanical stability in random loose packings can be
studied~\cite{Schroeter1}.
It is this regime of approximately homogeneous density in a stable
fluidized bed that we focus on here. We propose an event driven
algorithm to simulate three-dimensional fluidized beds. Our simulation
is based on a recently developed event-driven algorithm for hard
particles, experiencing drag by a surrounding
fluid~\cite{fiege2012dynamics,fiege2013anomalous}. Following Johri and Glasser
\cite{johri2002connections,johri2004bifurcation}, we assume a
\emph{density dependent} drag force and incorporate the density
dependence into the viscosity of the fluidizing liquid. The paper is
organized as follows: we first introduce the model (sec.~\ref{model})
and give details of the simulation (sec.~\ref{simulation}).
We then show that indeed an approximately homogeneous profile is
established (sec.~\ref{interface}). We determine the width of the
interface and finally discuss expansion and sedimentation after a
sudden change in density (sec.~\ref{sedimentation}).

\section{Model}\label{model}

We consider a monodisperse collection of $N$ spheres of mass $m$,
radius $R$ and volume $V= \frac{4}{3}\pi R^3$. Their density is
denoted by $\rho_s=m/V$. The particles are inserted in a fluid with
density $\rho_f$ and viscosity $\eta$. The fluid is flowing from below
to the top of the system with velocity $v_{ex}>0$. The particles are
subject to friction with the surrounding fluid, a gravitational force,
$g$, as well as bouyancy and inelastic collisions, so that the
Langevin equation for the velocity of particle $i$ in $z$-direction
reads
\begin{equation}
 \dot v_{i,z} = - \gamma(v_{i,z}-v_{ex}) -g + \frac{F_b}{m} + \left.\frac{dv_{i,z}}{dt}\right\vert_{coll}. \label{eq:langevin_fluidized}
\end{equation}
Here $ \phi = \frac{4}{3}\pi R^3 n_0$ denotes the packing fraction in
three dimensions and $n_0$ is the number density. The friction
coefficient in the drag force, $\gamma$, is allowed to depend on
volume fraction, $\gamma = \gamma_0 \cdot f(\phi)$, via the function
$f(\phi)$, which will be specified below. The bare friction
coefficient is given by the Stokes value: $\gamma_0=\frac{ 6 \eta \pi
  R}{m}$. We do not consider hydrodynamic interactions between the
granular particles and hence are restricted to dense suspensions,
where the hydrodynamic interactions are screened. The bouyancy force
$F_b$ is equal to the fluid mass that is displaced by the particle,
$F_b=\rho_f g V$.

The particles are modeled as hard particles which collide
inelastically. The degree of inelasticity is characterized by a
coefficient of incomplete restitution $\varepsilon$. Upon collision of
particles $i$ and $j$, their relative velocity $\mathbf g= \mathbf v_i
- \mathbf v_j$ changes to the postcollisional relative velocity
$\mathbf g^\prime$ according to
\begin{equation}
 (\mathbf g \cdot \mathbf n)^\prime =-\varepsilon (\mathbf g \cdot \mathbf n)
\end{equation}
where $\mathbf n = (\mathbf r_i -\mathbf r_j)/|\mathbf r_i -\mathbf
r_j| $. For the sake of simplicity, we assume a constant coefficent of
restitution $\varepsilon$ here.

The density dependence of $\gamma(\phi)$ is determined
experimentally~\cite{Schroeter4} by measuring the sedimentation
velocity, $U$, of the particles in a resting fluid. A single particle
in suspension moves with a constant velocity, denoted by $U_0$, when
gravity and bouyancy are balancing the frictional force.
With increasing $\phi$, the settling velocity $U$ decreases. Batchelor \cite{batchelor1972sedimentation} derived the first-order correction
\begin{equation}
 \frac{U}{U_0} = 1-6.55\phi + \mathcal{O}(\phi^2) \label{eq:expansion}\,.
\end{equation}
Higher order terms were derived in~\cite{cichocki2002three,cichocki2003three,batchelor2,batchelor3}, using a virial expansion.

The expansion Eq. (\ref{eq:expansion}) is helpful as long as dilute
packings are considered; in the case of dense fluids one must resort
to empirical relations.
Here we assume that the relation by \citet*{richardson1954u} captures
the settling behavior of the suspension appropriately as found in
recent experiments \cite{duru2002experimental,paetzold2012} and
simulations by \citet{abbas2006dynamics} and
\citet{yin2007hindered}. Hence we take
\begin{equation}
 \frac{U}{U_0} = (1-\phi)^n \label{eq:richardson}
\end{equation}
where $n$ depends on the Reynolds number.

Since we are interested in the density dependent damping
$\gamma(\phi)=f(\phi)\gamma_0$, we need a relation between $U/U_0$ and
$f(\phi)$. A sedimenting particle in a Newtonian liquid moves with
constant velocity $U_0$, when gravity, bouyancy and drag force balance,
$-F_D=F_G+F_B$, so that $-F_D=\frac{\gamma_0}{m}U_0
= (\rho_f-\rho_s)Vg$.  In a suspension, the sedimentation velocity
will depend on the packing fraction through the viscosity
$\gamma(\phi) =\gamma_0 f(\phi)$. Balancing forces then yields
$\frac{\gamma}{m}U=\frac{\gamma_0
  f(\phi)}{m}U=(\rho_f-\rho_s)Vg$. Dividing $U$ by $U_0$ we get
\begin{equation}
 \frac{U}{U_0}=\frac{1}{f(\phi)} \label{eq:u_u0_f_phi}
\end{equation}
so that we can obtain the density dependence of the damping
coefficient from the measured sedimentation velocities.  In the
following, we will use the Richardson-Zaki relation
(\ref{eq:richardson}) implying for $ \gamma(\phi)$
\begin{equation}
 \gamma(\phi) = \gamma_0 \cdot (1-\phi)^{-n} \,.
\end{equation}




In the stationary state, the average particle velocity
vanishes, $\sum_i v_{i,z}=0$, and the external flow field, $v_{ex}$, controls the average packing fraction, $\bar\phi$ according to 
\begin{equation}
 \dot v_{i,z}=0=\gamma(\bar\phi)\bar v_{ex} -g  \frac{\rho_s - \rho_f}{\rho_s}. 
 \label{eq:homog_solution}
\end{equation}
We now expand Eq.(\ref{eq:langevin_fluidized}) around the homogeneous
stationary state and keep only linear deviations: $\phi(\mathbf
r_i)=\bar\phi+\delta\phi_i$ and $\gamma = \gamma(\bar\phi) +
\gamma^\prime (\bar\phi ) \delta \phi_i$. We also allow for
fluctuations in the external flow field $v_{ex}\to v_{ex} + \xi_i$.
Linearizing Eq. (\ref{eq:langevin_fluidized}) around the stationary state
of Eq. (\ref{eq:homog_solution}) we find
\begin{equation}
\begin{split}
  \dot v_{i,z}=  - \gamma(\bar\phi)v_{i,z} +  \gamma(\bar\phi)\xi_{i,z} +  \left.\frac{dv_{i,z}}{dt}\right\vert_{coll} \\  +\frac{\gamma^\prime(\bar\phi)}{\gamma(\bar\phi)}  \frac{\rho_s - \rho_f}{\rho_s} g \delta \phi_i 
\end{split}
\label{eq:langevin_new}\end{equation}
The driving
force towards the stationary state in Eq.(\ref{eq:langevin_new}) is
the density, or more precisely its deviation from the average
value. If locally $\delta \phi_i<0$, particles in this region will be
accelerated downward, if $\delta \phi_i>0$, particles will be
accelerated upward. The equations of motion (\ref{eq:langevin_new})
can be simulated with an event driven algorithm, as detailed in the
next paragraph.

\section{Event driven simulation}
\label{simulation}


To implement Eq.~(\ref{eq:langevin_new}) in an event driven algorithm,
we model the driving force, resulting from the inbalance of gravity
and buoyancy, by discrete \emph{kicks}. In other words, the particles
are not accelerated continuously, but instead are kicked with a given
frequency. These kicks are treated as events in the simulation, with
the kick frequency, $f_{dr}$, comparable to the collision frequency,
$f_{coll}$.

Similarly, the noise $\{\xi_{i,\alpha}\}$ is modelled by \emph{random}
kicks with average zero and variance $\xi_0^2$. In the stationary
state, energy losses due to drag forces and inelastic collisions
balance energy input due to driving, according to:
\begin{equation}
\frac{3}{4} f_{coll} (1-\varepsilon^2) T_G + 
2\gamma(\bar\phi) T_G= f_{dr} m  \gamma^2(\bar\phi)\xi_0^2.\nonumber
\end{equation}
Here we have introduced the granular temperature $T_G=2/3 E_{kin}$ in
terms of the average kinetic energy (for details see
\cite{fiege2012dynamics}). 

In experiments on fluidized beds of granular particles, the thermal
heat bath provided by the surrounding fluid is small in the sense that the
gravitational energy of a grain is much larger than the thermal
energy. Nevertheless the grains show random motion with typical
velocities of the order of a few mm/s. Energy is fed into the system by
pushing a liquid through the fluidized bed. This energy input does not
only compensate the frictional losses due to drag
but in
addition provides the energy input to sustain the random motion of the
particles in the stationary state.

We are interested in the fluidization and sedimentation of the
system. The density dependent contribution to the particle motion in
$x$- and $y$-direction is therefore negligible as also found by
\citet{nguyen2005sedimentation}. To compute the local packing fraction
$\phi_i$, we count the number of particles in a layer of defined
thickness $\Delta$ around each particle at the time of the kick,
which determines the deviation of the local density from its global
value, $\delta \phi_i =\phi_i - \bar\phi $.  For most of the
simulations we choose $\Delta=10R$, but have checked other values (see
below).

Finally, we use the Richardson-Zaki relation (\ref{eq:richardson}) to
rewrite Eq.~(\ref{eq:langevin_new})
\begin{equation}
 \dot v_{i,\alpha} = -\gamma(\bar\phi)(v_{i,\alpha}-\xi_{i,\alpha}) + \left.\frac{dv_{i,\alpha}}{dt}\right\vert_{coll}+ \delta_{z,\alpha}b \frac{\delta \phi_i}{1-\bar\phi}
\label{eq:sim_fl_bed}
\end{equation}
with $b=ng(\rho_s-\rho_f)/\rho_s>0$.

We use systems of $N=10 000 $ and $N=200 000$ monodisperse particles
in a box with hard walls, i.e. colliding particles are reflected
elastically. At high densities, we have to circumvent the inelastic
collapse, which is done in the same way as in \cite{fiege2009long}.
We regard our ansatz as a simplification for more complicated
simulations as in \cite{xu1997numerical} and
\cite{hoomans2000granular}, where $2400$ and $5000$ particles,
respectively, were simulated. Both methods are combinations of
molecular dynamics, capturing the motion of single particles, and
computational fluid dynamics, accounting for the interstitial
fluid. The main advantage of our simulation ansatz is that the
event-driven simulation can handle $N=\mathcal{O}(10^6)$ and more
particles.

{\it Choice of parameters}: The range of packing fractions is chosen to be
$0.3\leq\bar\phi\leq0.55$ and $n=4.65$ as discussed in
\cite{paetzold2012}. We choose the parameters as $\varepsilon=0.7$,
and $\frac{\rho_s-\rho_f}{\rho_s} = 0.6$ , to mimick a typical
experiment with glass spheres. The drag coefficient is determined from
Stokes law $\gamma_0=6\pi\eta R/m$, with the viscosity of water
$\eta\sim 0.01$g cm/s. An important parameter is the diameter $d=2R$
of the particles, because it determines the ratio, $Z$, of the
potential energy gain by lifting a particle by its diameter as
compared to the thermal energy
\begin{equation}
Z=\frac{\pi\Delta\rho g d^4}{3 k_B T}
\end{equation}
with $\Delta\rho=\rho_s-\rho_f$. For a typical example of an
experimental realisation, we choose $R=600\mu m$, mass $m= \rho_s
\cdot V = 2.262\cdot 10^{-6}$kg and $\gamma_0=5/$s. The granular
temperature of the undisturbed system (no flow, no gravity) is set to
$k_BT=2\cdot 10^{-9}\text{kg m}^2/\text{s}^2$ corresponding to a typical velocity of mm/s.
For these parameters $Z\sim 16$.

In the results section, we will use these specific values with the
experiments of Ref.~\cite{paetzold2012} in mind. However, we want to point out
that the same data also apply to other systems: Since we model the
particles as hard spheres, there is no inherent length scale
associated with the interaction and the particle diameter basically
sets the unit of length. In other words the same data can also be
interpreted in terms of differently sized particles. For example
choosing $R=3$mm corresponds to a mass
$m=\rho_s\frac{4\pi}{3}R^3=2.8 10^{-4}$kg and a drag coefficient
$\gamma_0=0.2/$s. If the noise level is kept constant these parameters
imply a granular temperature $k_BT=10^{-8}\text{kg m}^2/\text{s}^2$ and $Z\sim
2000$.

\section{Results}

Our primary interest is the fluidization and sedimentation in a
granular suspension, driven by an upward flow. As a first step we
compute the density profile for a given average volume fraction,
corresponding to a given flow rate~(\ref{eq:u_u0_f_phi}). We characterize the
resulting interface and check for effects due to the finite resolution
for the density $\delta \phi_i$. Subsequently the dynamics of
fluidization and sedimentation is simulated by changing the average
volume fraction and monitoring the following expansion or
compactification. 

\subsection{Interface formation and packing fraction profile}
\label{interface}

As a first step, we investigate the formation of an interface for a
given average packing fraction $\bar\phi$. To that end we prepare the
system initially in a maximally homogeneous state by equilibrating the
system without gravity in a closed box, corresponding to the desired
volume fraction. Subsequently the simulation box is enlarged in the
z-direction and gravity and density dependent drag force are switched
on.

\begin{figure}
 \begin{center}
 \includegraphics[width=0.5\textwidth]{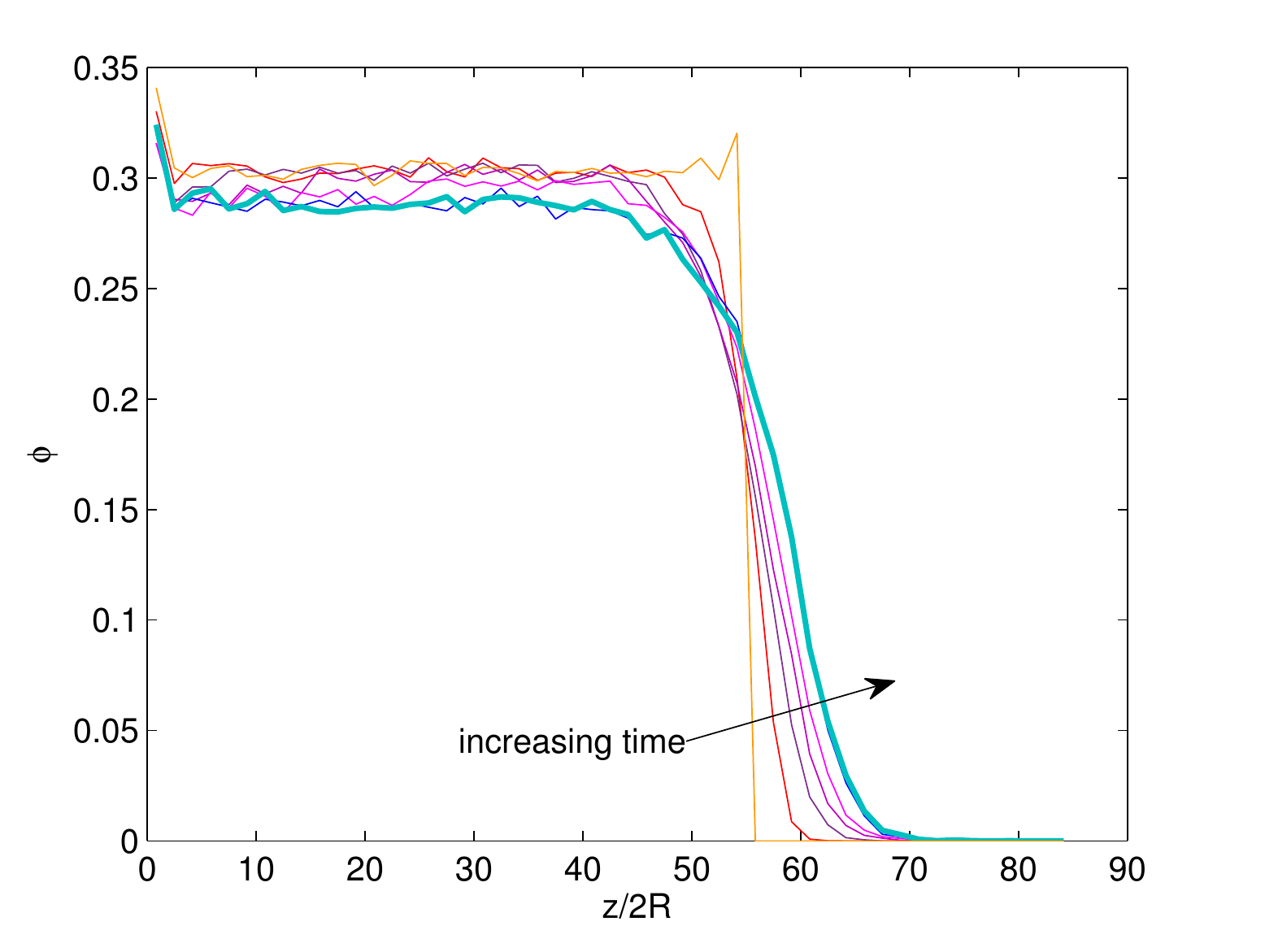}
\end{center}
\caption{Interface formation for a fluidized bed of packing fraction
  $\bar\phi=0.3$. Time step in between two subsequent density profiles
  $\sim 20,50,100,200,400,400$ collisions per particle.}
\label{fig:interface}
\end{figure}

We monitor the particles' position to compute the density profile
\begin{equation}
 \phi(z)=\frac{\bar\phi}{N}\sum_i \delta(z-z_i)
\end{equation}
In Fig.~\ref{fig:interface} this profile is shown at different times,
starting from a sharp profile. We observe the formation of an
interface of finite width with a stationary state reached after around
$1000$ collisions per particle. At the bottom $z\approx R $, the
packing fraction is locally
increased, 
because the particles tend to arrange at a height $z\approx R$ as this
configuration allows for the most efficient packing.

A simple model can account for the observed profiles and allows us to
quantify the width of the interface. We consider a bed, extended to
negative $z$, so that the probability to find particle $i$ at position
$z$ in the homogeneous state is given by
\begin{equation}
p_{hom}(z) =p_0\theta(h-z)
 \label{eq:homogprob}
\end{equation}
At the interface, several processes disturb the particle position. We
model these by a  Gaussian with zero mean and variance $\sigma^2$.
Hence the probability to find
the particle at position $z$ is given by a convolution
\begin{eqnarray}
 p(z)& =& (p_{hom} \star p_\sigma )(z) \nonumber \\ 
 & =&\frac{p_0}{2}\left(1 -\textrm{erf} \left( \frac{z-h}{\sqrt{2}\sigma} \right)  \right)\,.\label{eq:convolutionflbed}
\end{eqnarray}
The packing fraction is proportional to this probability and the proportionality constant can be fixed by requiring $\phi=\bar\phi$ well inisde the bulk 
\begin{equation}
 \phi(z)= \frac{\bar\phi}{2}\left(\textrm{erfc} \left( \frac{z-h}{\sqrt{2}\sigma} \right)  \right) \,.\label{eq:convolutionflbed}
\end{equation}
Hence we have two free parameters, we call $h$ the height of the bed and $\sigma$ is a measure for the sharpness of the interface: The smaller $\sigma$ the sharper the interface. 



\begin{figure}
 \begin{center}
 \includegraphics[width=0.5\textwidth]{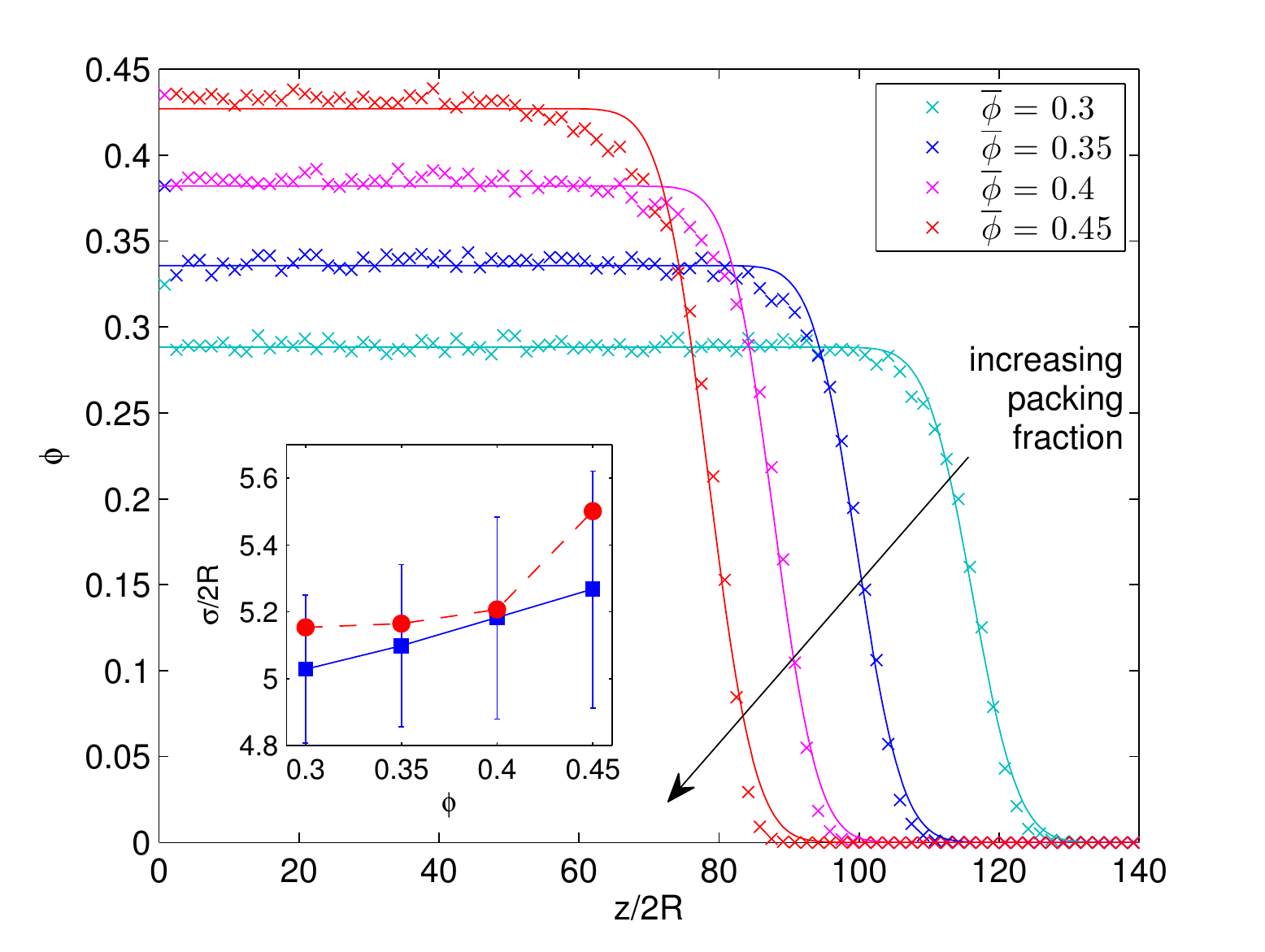}
\end{center}
\caption{Packing fraction profiles for fluidized states of different
  flow rates, corresponding to stationary packing fractions
  $\bar\phi=0.3$. $0.35$, $0.4$ and $0.45$. Data points (crosses) are
  fitted (solid lines) to Eq. \ref{eq:convolutionflbed}; inset: width
  of the interface as a function of average packing fraction
  $\bar\phi$ for two different values of $\Delta$.}
\label{fig:interfacefit}
\end{figure}


We fit the packing fraction profile to Eq.(\ref{eq:convolutionflbed})
for fluidized states of several packing fractions. As discussed above,
at the system bottom the packing fraction is locally
increased. Therefore, we exclude these data points from the following
procedure. The data and the resulting fits are plotted in
Fig. \ref{fig:interfacefit}. Our modeled packing fraction profile in
Eq. (\ref{eq:convolutionflbed}) matches the measured data very well
for moderately dense packings, but deviates for higher densities in
the region where the bulk value starts to drop down. Nevertheless we
can use the fit procedure to determine the width of the interface,
which is shown in the inset of Fig.~\ref{fig:interfacefit} as a
function of packing fraction. The width of the interface $\sigma$ is
in the range of $10$ particle radii and is approximately
independent of $\phi$.


To calculate the local packing fraction we have used a slice of
thickness $\Delta= 10R$. We do not expect that the resolution in the
local density has strong effects in the bulk of the sample. However
the interface and in particular its width might depend on $\Delta$. To
check this point we simulate the same fluidized bed but with a
different value of $\Delta = 6R$. The width of the interface $\sigma$
is compared for the two resolutions in the inset of
Fig. \ref{fig:interfacefit}. On average, $\sigma$ is slightly smaller
for $\Delta=6R$. However the effect is small and we do not expect that
a further refinement of the resolution for the density will change
these results.

\subsection{Sedimentation and Expansion}
\label{sedimentation}
\begin{figure}[h]
 \begin{center}
 \includegraphics[width=.48\textwidth]{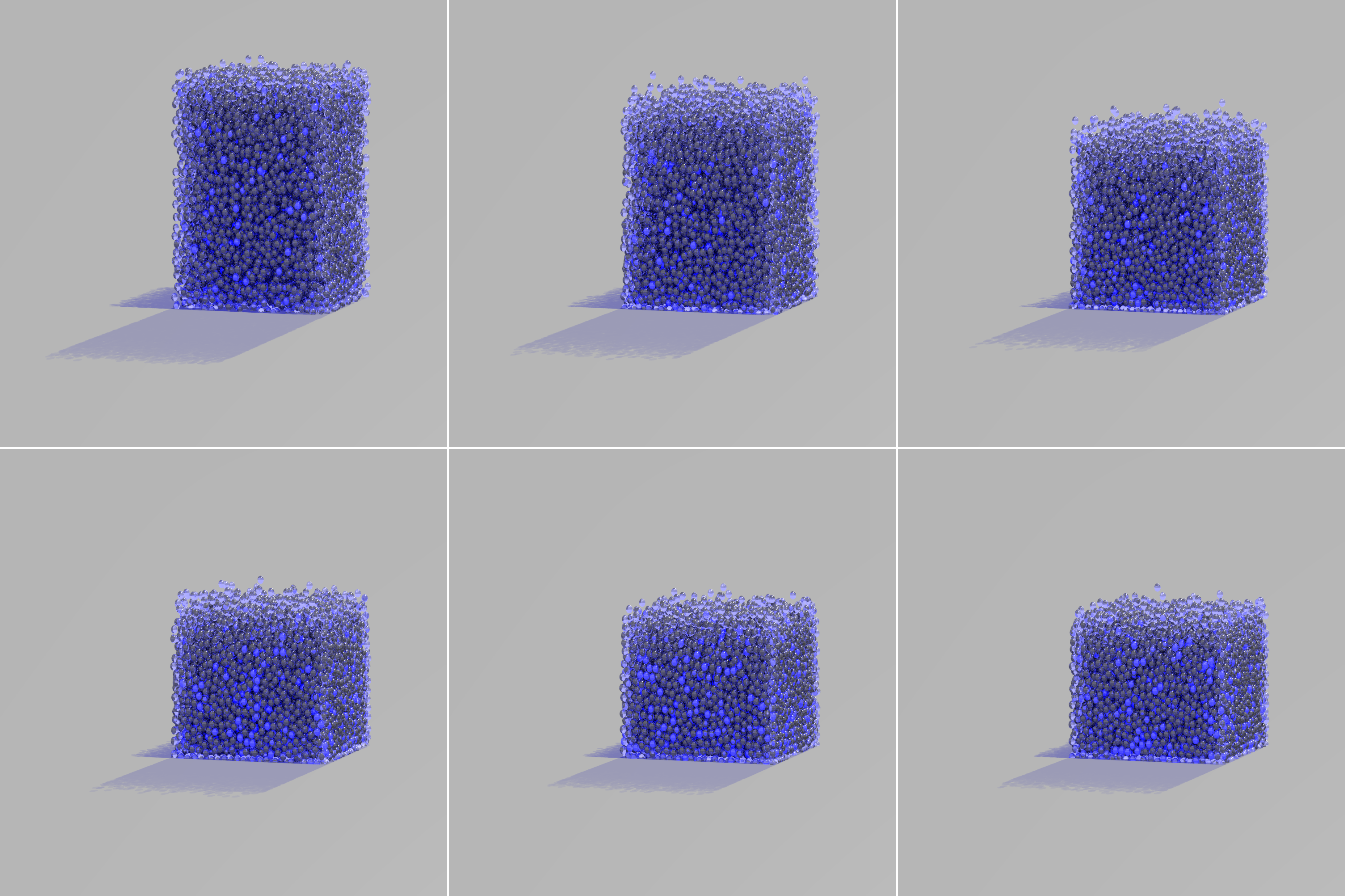}
\end{center}
\caption{Sedimentation process from $\bar\phi=0.3$ to $\bar\phi=0.55$,
  at times $t= 0,0.4,0.8,1.2,1.6,2.0 s$. At the upper left, the bulk
  has a packing fraction $\approx 0.3$, at the lower right $\approx
  0.55$, see also Fig. \ref{fig:sedimentationprocess}.}
\label{fig:sedimentation}
\end{figure}

So far we have shown that our simple model allows for a stationary
state with a well defined interface whose width has been
characterized. In this section we study the relaxational dynamics of
the fluidization and sedimentation process which in experiment is
achieved by changing the flow rate. The initial and final flow rate
give rise to different stationary packing fractions. In our
simple model sedimentation and compactification is modeled by a sudden
(instantaneous) change in density. Specifically, we compact a
fluidized bed with $N=10^4$ particles from $\bar\phi=0.3$ to a
different packing fraction $\bar\phi =0.55$. The initial state is
chosen as the stationary state for $\bar\phi=0.3$ and then at $t=0$
the average volume fraction is set to $\bar\phi =0.55$ in
Eq.(\ref{eq:sim_fl_bed}). To illustrate the sedimentation process, we
depict the compactifying bed at six different times during the
sedimentation in Fig. \ref{fig:sedimentation}. Already from this
figure, especially the second and third frame, one can see that the
sedimentation process is not homogeneous.

\begin{figure}
 \begin{center}
 \includegraphics[width=0.5\textwidth]{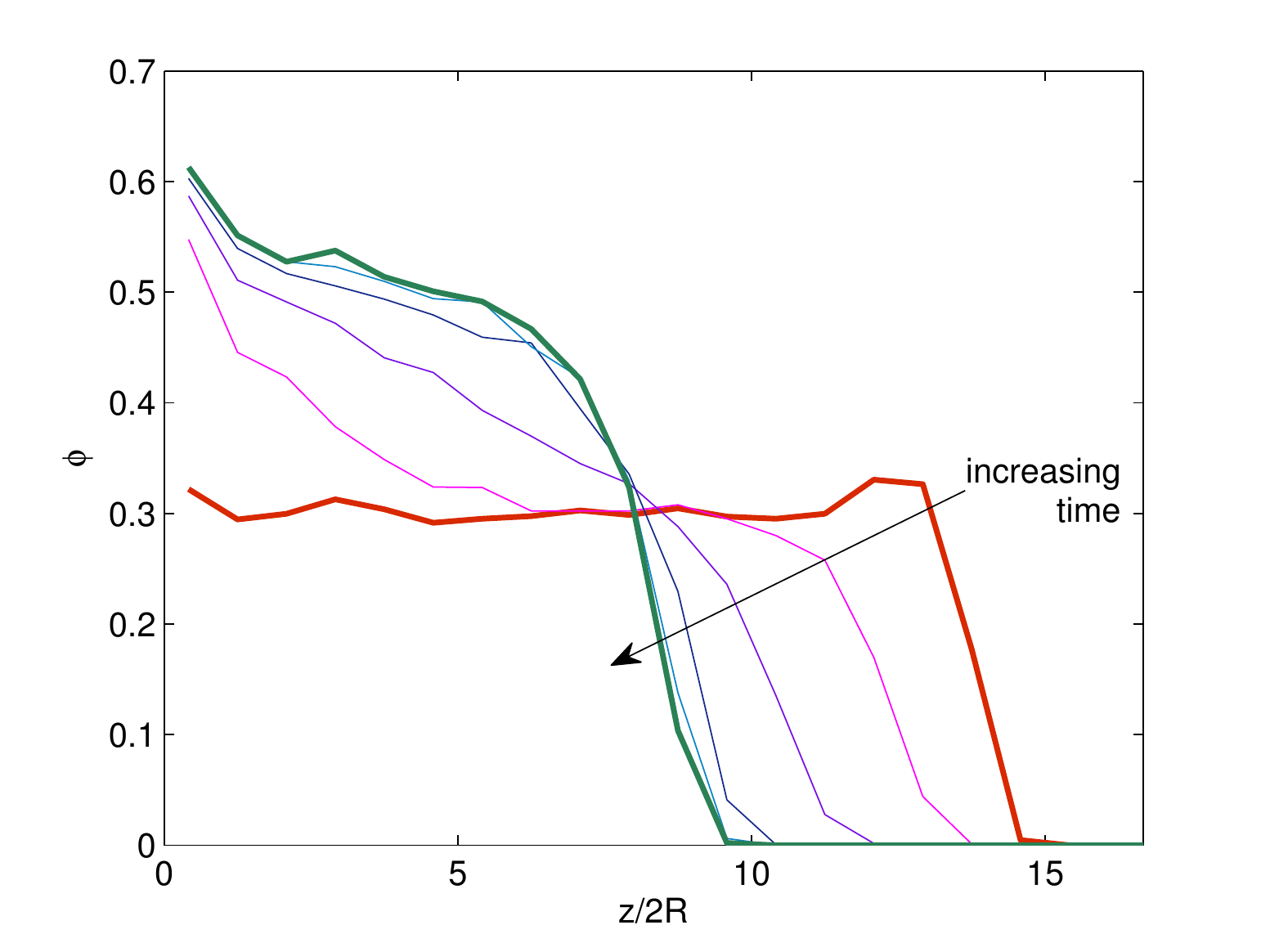}
\end{center}
\caption{Packing fraction profiles for sedimentation from
  $\bar\phi=0.3$ to $\bar\phi=0.55$, time step $\Delta t = 0.4s$. Same
  data as in Fig. \ref{fig:sedimentation}.}
\label{fig:sedimentationprocess}
\end{figure}

To quantify this process, we plot in
Fig.\ref{fig:sedimentationprocess} the density profile for several
intermediate time steps. The compactification starts from the bottom
of the fluidized bed, working its way up to the top. The packing
fraction starts to increase to the final value $\bar\phi=0.55$ at the
bottom, subsequently the layers above are compacting and after about
$500$ collisions per particle at $t=2s$ the new stationary state is
attained.

\begin{figure}
 \begin{center}
 \includegraphics[width=0.5\textwidth]{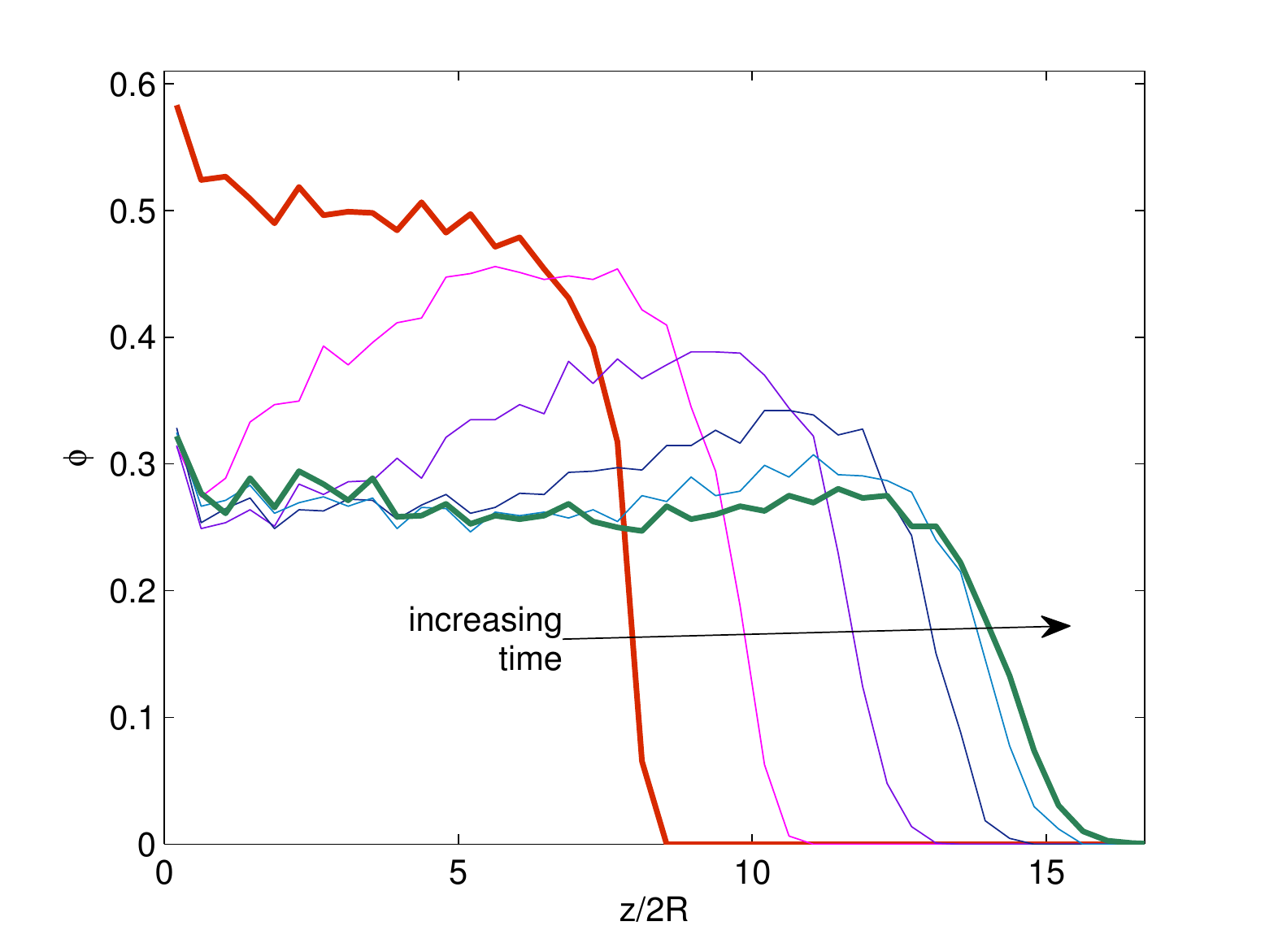}
\end{center}
\caption{Expansion from $\bar\phi=0.55$ to $\bar\phi=0.3$, time step
  $\Delta t = 0.16$.}
\label{fig:fluidizationprocess}
\end{figure}

Similarly, we monitor the reverse process, i.e. expansion of the
bed. The initial state is chosen as the stationary state for
$\bar\phi=0.55$ and then at $t=0$ the average volume fraction is set
to $\bar\phi =0.3$ in Eq.(\ref{eq:sim_fl_bed}). The density profile is
shown in Fig.\ref{fig:fluidizationprocess} for several timesteps.
The expansion process is inhomogeneous as well. The bed rapidly
expands at the top with a simultaneous inhomogeneous dilution in the
compactified region. Subsequently the upper part of the system
continues to expand until the packing fraction has ultimately
flattened throughout the system. The expansion process is faster
than the sedimentation. After about $90$ collisions per particle at
time $t=0.8s$ the bed has attained the expanded state.

\begin{figure}
 \begin{center}
 \includegraphics[width=0.5\textwidth]{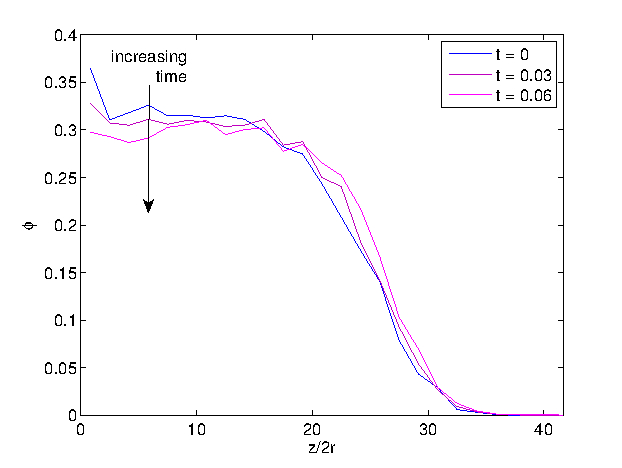}
\end{center}
\caption{Expansion from $\bar\phi=0.35$ to $\bar\phi=0.3$.}
\label{fig:slow_expansion}
\end{figure} 
The above expansion corresponds to a drop of almost $50\%$ in density
and is by no means quasistatic. To study the latter, we consider a
much smaller drop in density, from $\bar\phi=0.35 \to 0.3$. The
profiles are shown in Fig.~\ref{fig:slow_expansion} and observed to be
approximately monotonic as a function of height within the bed.

\begin{figure}
 \begin{center}
 \includegraphics[width=0.5\textwidth]{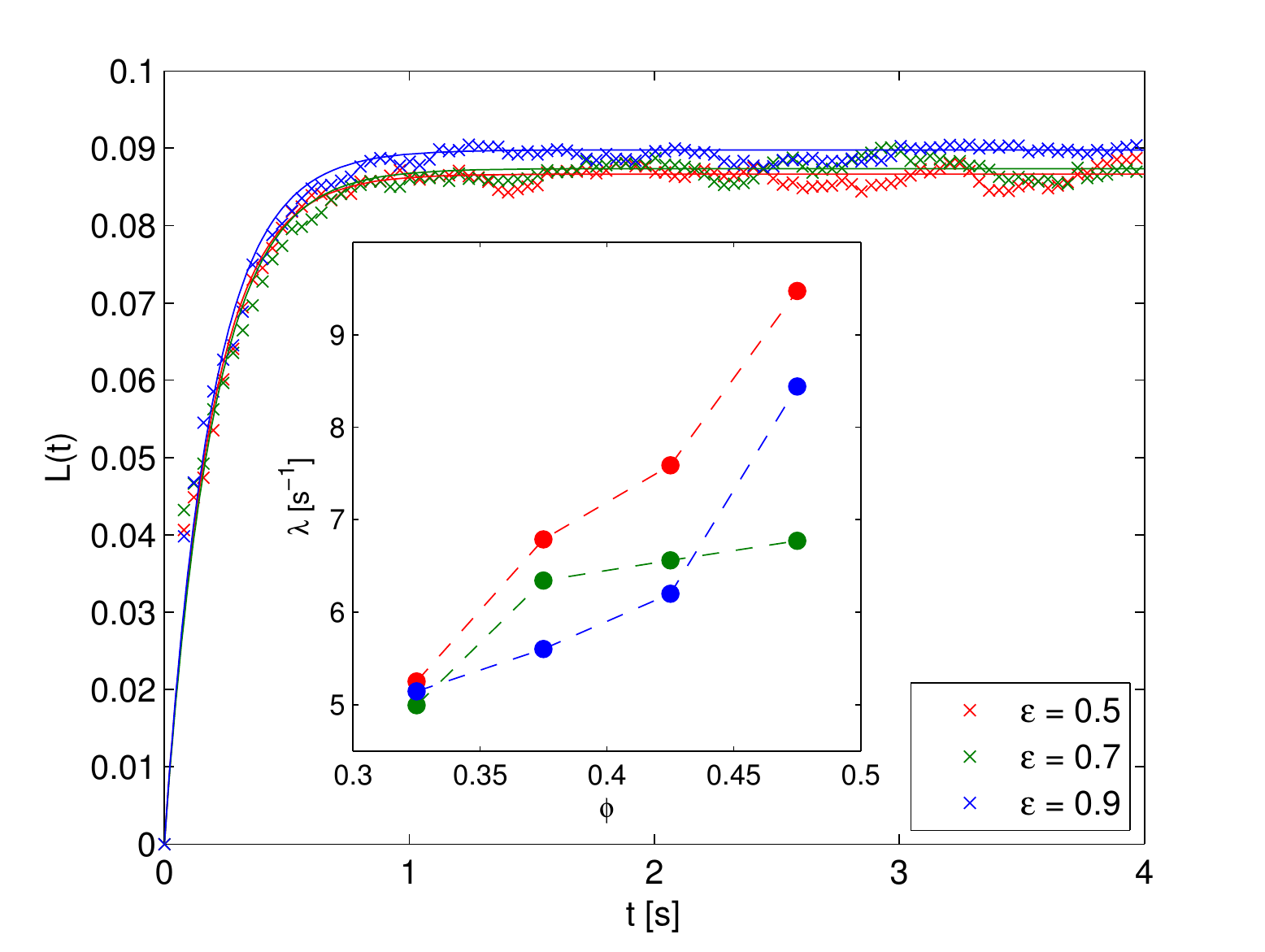}
\end{center}
\caption{Approach to the stationary state, as measured by $L(t)$ for the
expansion from $\bar\phi=0.35$ to $\bar\phi=0.3$; inset: $\lambda(\bar\phi)$ for three different $\varepsilon$}
\label{fig:Lt}
\end{figure}

We want to extract a timescale for the expansion and study its
dependence on average packing fraction and coefficient of restitution.
To that end we define the average deviation from the initial profile
\begin{equation}
L^2(t)=\int_0^{\infty}dz|\phi(z,t)-\phi(z,0)|^2
\end{equation}
and compute its temporal evolution. As seen in Fig.~\ref{fig:Lt},
there is a rapid increase for short times followed by a plateau at
longer times. The data are well fitted by an exponential increase
according to: $L(t)=const. (1-e^{-\lambda t})$. In the inset of
Fig.~\ref{fig:Lt} we show $\lambda(\bar\phi)$ as a function of average
packing fraction for small changes of packing fraction, i.e. the first
point corresponds to $\bar\phi=0.35$ to $\bar\phi=0.3$, the second
point to $\bar\phi=0.4$ to $\bar\phi=0.35$ and so on. The timescale of
expansion increases with increasing volume fraction and is slightly
larger for the more inelastic systems.

\section{Conclusion and Outlook}

We have shown how to simulate a fluidized bed with an event-driven
method that is capable of simulating large numbers of particles. The
algorithm is based on an expansion around the homogeneous fluidized
state.

We found stable homogeneous fluidized beds, whose packing fraction can
be adjusted by changing the flow rate. A stable bed is characterized
by an interface at the top of the system, whose width is of the order
of 5-10 particle radii. A stable bed also forms after changing the
flow rate to a lower value, which increases the packing fraction
(sedimentation), and by increasing the flow rate, which decreases the
packing fraction (expansion). In general, the processes of
sedimentation and expansion are non homogeneous, i.e. the packing
fraction in the dense region does not deform homogeneously but rather
changes from bottom to top, ultimately flattening to the previously
observed stationary profiles.

We plan to further investigate fluidized beds using our algorithm. An
obvious first extension is the temperature profile. Furthermore, we
can easily compute velocity distributions and mean square
displacements, which are also accessible to experiment by introducing
markers.

Here, we have focused on approximately homogeneous states. However, it
is known~\cite{Guazzelli_2011} that suspensions of sedimenting
particles exhibit a rich spectrum of instabilities. We have already
seen striped phases in our simulations for very small noise levels,
but postpone a sytematic study to future work.









\begin{acknowledgments}
  We thank W. T. Kranz, M. Schroeter and S. Ulrich for useful
  discussions. We furthermore acknowledge support from DFG by FOR
  1394.
\end{acknowledgments}
\bibliography{lit2.bib}

\end{document}